# Magnetoplasma waves on the semiconductor nanotube surface

**A.M. Ermolaev and G.I. Rashba**

Department of Theoretical Physics, V.N. Karazin Kharkiv National University,

4 Svobody square, 61077, Kharkiv, Ukraine

E-mail: georgiy.i.rashba@univer.kharkov.ua

## Abstract

The effective mass approximation is used to consider plasma and magnetoplasma waves in an electron system on the surface of the semiconductor cylindrical nanotube. The electron-electron coupling is taken into account in the random phase approach. In the case of a degenerate electron gas the spectral windows on the wavevector-frequency plane and the spectra of the waves are obtained. Their frequencies undergo quantum oscillations of the de Haas-van Alfven type which are attributed to the Fermi level traversing the sub-zone boundaries in the electron energy spectrum. The spectrum and the damping of waves in the non-degenerate electron gas were found. In a magnetic field parallel to the cylinder axis the frequencies of the magnetoplasma waves sustain the Aharonov-Bohm type oscillations that appear with changing magnetic field strength.

PACS numbers: 73.22.Lp

## Introduction

The growing interest of researchers in electron nanosystems on curving surfaces [1,2] can be attributed to a diverse number of reasons. These systems are functional elements of many devices and engineering gadgets. The present perfection of the experimental setups allow for production of these systems in a laboratory framework. For theorists these systems are convenient objects for exploring novel methods of computations of physical values. The curvature of the structure and external magnetic field enrich the picture of phenomena occurring in nanostructures. The methods to control their properties become more diverse.

Various effects have been found on curving surfaces in such electron systems, which can not be reproduced in the plane geometry case. Among them, are the effects of hybridization of the spatial and magnetic quantization of electron motion, the modification of the hamiltonian of the electron system [2,3], the unusual performance of conductance [4] and of magnetic response of the system [5], the peculiarities of the shielding of electron-electron coupling [6], the specific resonances in electron scattering, produced by impurity atoms in carbon nanotubes [7] and quantum wires [8], etc.

The typical examples of the curving surface nanosystems are semiconductor nanotubes [1-7]. The electron energy spectrum in these systems has a zone structure. It allows, due to a small number of electrons near the zone bottom, the use of the effective mass approximation.

This approximation enables to describe qualitatively, and often quantitatively, the properties of this kind of systems.

Plasma oscillations of the electron gas density on the curving surface were investigated in Ref. [9]. However, the dispersion relationships of the spectrum and damping of plasmons were not given [9]. The present paper considers plasma and magnetoplasma waves on the surface of a semiconductor cylindrical nanotube. We use the effective mass approximation, with the electron-electron coupling taken into account in the random phase approach. We consider the spectrum of density oscillations of the degenerate electron gas in Section 1 and the non-degenerate gas in Section 2. The magnetoplasma waves on the cylinder surface in the longitudinal magnetic field are considered in Section 3.

## I. Degenerate electron gas

In the effective mass approximation the wave function of a stationary electron state on the surface of semiconductor cylindrical nanotube has the following form:

$$\psi_{mk}(\varphi, z) = \frac{1}{\sqrt{2\pi}} e^{im\varphi} \frac{1}{\sqrt{L}} e^{ikz}, \qquad (1)$$

where $m = 0, \pm 1, ...$ is the azimuthal quantum number, $k$ is the projection of electron wave vector to the cylinder axis $z$, $\varphi$ is the polar angle, $L$ is the length of the tube. The electron energy in the state (1) is:

$$\varepsilon_{mk} = \varepsilon_0 m^2 + \frac{k^2}{2m_*}. \qquad (2)$$

Here $m_*$ is the effective electron mass, $\varepsilon_0 = 1/2m_* a^2$ is the gyration quantum, $a$ is the tube radius. Herein the quantum constant is assumed to be equal to unity.

The spectrum (2) consists of many one-dimensional sub-zones. The electron state density,

$$\nu(\varepsilon) = \frac{\sqrt{2m_*} L}{\pi} \sum_m \frac{1}{\sqrt{\varepsilon - \varepsilon_0 m^2}}, \qquad (3)$$

contains square root singularities at the sub-zone boundaries, $\varepsilon_m = \varepsilon_0 m^2$. In this formula the summation is done over those values of $m$ for which the expression under the root is positive. Calculating the sum according to the Poisson formula we obtain:

$$\nu(\varepsilon) = 2m_* aL \left[ 1 + 2 \sum_{l=1}^{\infty} J_0 \left( 2\pi l \sqrt{\frac{\varepsilon}{\varepsilon_0}} \right) \right],$$

where $J_0$ is the Bessel function. In the limit $\varepsilon \gg \varepsilon_0$, we have

$$\nu(\varepsilon) = 2m_* aL \left[ 1 + \frac{2}{\pi} \left( \frac{\varepsilon_0}{\varepsilon} \right)^{1/4} \sum_{l=1}^{\infty} \frac{1}{\sqrt{l}} \cos\left( 2\pi l \sqrt{\frac{\varepsilon}{\varepsilon_0}} - \frac{\pi}{4} \right) \right].$$

The density of states oscillates with varying $\sqrt{\varepsilon}$ with the period $(\sqrt{2m_*} a)^{-1}$. The monotone term $2m_* aL$ is equal to the density of states of the two-dimensional electron gas on the area $S = 2\pi aL$.

In the random phase approximation, the dispersion equation for the spectrum of plasma waves on the surface of the nanotube has the form [6]:

$$1 - \frac{\upsilon_m(q)}{2\pi L} P_m(q,\omega) = 0, \tag{4}$$

where

$$\upsilon_m(q) = 4\pi \bar{e}^2 I_m(a|q|) K_m(a|q|) \tag{5}$$

is the cylindrical harmonic of the electron Coulomb interaction energy, $\bar{e}$ is the electron charge divided by the background dielectric constant, $I_m$ and $K_m$ are the modified Bessel functions [10], $P_m$ is the delayed polarization operator which depends on the projection $q$ of the wave vector to the axis $z$ and frequency $\omega$. With the aim to consider further the magentoplasma waves, we shall derive the value of $P_m$ for the electron gas in the magnetic field $B$, which is parallel to the cylinder axis,

$$P_m(q,\omega) = \sum_{m'k\sigma} \frac{f(\varepsilon_{(m'+m)(k+q)\sigma}) - f(\varepsilon_{m'k\sigma})}{\varepsilon_{(m'+m)(k+q)\sigma} - \varepsilon_{m'k\sigma} - \omega - i0}, \tag{6}$$

where

$$\varepsilon_{mk\sigma} = \varepsilon_0 \left(m + \frac{\Phi}{\Phi_0}\right)^2 + \frac{k^2}{2m_*} + \sigma\mu_B B \tag{7}$$

is the electron energy in magnetic field [5], $\Phi = \pi a^2 B$ is the magnetic flux through the tube cross-section, $\Phi_0 = 2\pi c/e$ is the flux quantum ($e$ is the electron charge, $c$ is the speed of light), $\mu_B$ is the electron spin magnetic moment, $\sigma = \pm 1$ is the spin quantum number, $f$ is the Fermi function.

The function (6) depends on $|q|$ and satisfies the following symmetry properties:

$$\begin{aligned}
\operatorname{Re} P_m(q,-\omega,-\Phi) &= \operatorname{Re} P_m(q,\omega,\Phi), \\
\operatorname{Im} P_m(q,-\omega,-\Phi) &= -\operatorname{Im} P_m(q,\omega,\Phi), \\
P_{-m}(q,\omega,-\Phi) &= P_m(q,\omega,\Phi).
\end{aligned} \tag{8}$$

Using $P_m = \sum_{m'\sigma} P_{mm'}^{\sigma}$, we obtain at zero temperature:

$$\operatorname{Re} P_{mm'}^{\sigma}(q,\omega) = -\frac{m_* L}{2\pi q} \left( \ln\left| \frac{-q\upsilon_{m'+m}^{\sigma} - \omega_q + \Omega_{mm'} - \omega}{-q\upsilon_{m'}^{\sigma} + \omega_q + \Omega_{mm'} - \omega} \right| + \ln\left| \frac{q\upsilon_{m'}^{\sigma} + \omega_q + \Omega_{mm'} - \omega}{q\upsilon_{m'+m}^{\sigma} - \omega_q + \Omega_{mm'} - \omega} \right| \right),$$

$$0 < q < k_{m'}^{\sigma} - k_{m'+m}^{\sigma};$$

$$\operatorname{Re} P_{mm'}^{\sigma}(q,\omega) = -\frac{m_* L}{2\pi q} \ln\left| \frac{q\upsilon_{m'}^{\sigma} + \omega_q + \Omega_{mm'} - \omega}{q\upsilon_{m'+m}^{\sigma} - \omega_q + \Omega_{mm'} - \omega} \right|, \tag{9}$$

$$k_{m'}^{\sigma} - k_{m'+m}^{\sigma} < q < k_{m'}^{\sigma} + k_{m'+m}^{\sigma};$$

$$\operatorname{Re} P_{mm'}^{\sigma}(q,\omega) = -\frac{m_* L}{2\pi q} \ln\left| \frac{q\upsilon_{m'}^{\sigma} + \omega_q + \Omega_{mm'} - \omega}{-q\upsilon_{m'}^{\sigma} + \omega_q + \Omega_{mm'} - \omega} \right|,$$

$$q > k_{m'}^{\sigma} + k_{m'+m}^{\sigma};$$

$$\mathrm{Im}\, P^{\sigma}_{mm'}(q,\omega) = -\frac{m_* L}{2|q|}\Big\{\Theta\big[\omega-\big(-q\upsilon^{\sigma}_{m'}+\omega_q+\Omega_{mm'}\big)\big]\Theta\big[-q\upsilon^{\sigma}_{m'+m}-\omega_q+\Omega_{mm'}-\omega\big]+$$

$$+\Theta\big[\omega-\big(q\upsilon^{\sigma}_{m'+m}-\omega_q+\Omega_{mm'}\big)\big]\Theta\big[q\upsilon^{\sigma}_{m'}+\omega_q+\Omega_{mm'}-\omega\big]\Big\},$$

$$0 < q < k^{\sigma}_{m'}-k^{\sigma}_{m'+m};$$

$$\mathrm{Im}\, P^{\sigma}_{mm'}(q,\omega) = -\frac{m_* L}{2|q|}\Theta\big[\omega-\big(q\upsilon^{\sigma}_{m'+m}-\omega_q+\Omega_{mm'}\big)\big]\Theta\big[q\upsilon^{\sigma}_{m'}+\omega_q+\Omega_{mm'}-\omega\big], \qquad (10)$$

$$k^{\sigma}_{m'}-k^{\sigma}_{m'+m} < q < k^{\sigma}_{m'}+k^{\sigma}_{m'+m};$$

$$\mathrm{Im}\, P^{\sigma}_{mm'}(q,\omega) = -\frac{m_* L}{2|q|}\Theta\big[\omega-\big(-q\upsilon^{\sigma}_{m'}+\omega_q+\Omega_{mm'}\big)\big]\Theta\big[q\upsilon^{\sigma}_{m'}+\omega_q+\Omega_{mm'}-\omega\big],$$

$$q > k^{\sigma}_{m'}+k^{\sigma}_{m'+m}.$$

Here

$$\upsilon^{\sigma}_m = \frac{k^{\sigma}_m}{m_*} = \sqrt{\frac{2}{m_*}}\sqrt{\mu_0-\varepsilon_{m\sigma}} \qquad (11)$$

is the electron limiting velocity in the sub-zone with the number $(m,\sigma)$,

$$\varepsilon_{m\sigma} = \varepsilon_0\left(m+\frac{\Phi}{\Phi_0}\right)^2 + \sigma\mu_B B$$

is the sub-zone boundary, $\omega_q = q^2/2m_*$, $\mu_0$ is the chemical potential at zero temperature,

$$\Omega_{mm'} = \varepsilon_{(m'+m)\sigma}-\varepsilon_{m'\sigma} \qquad (12)$$

are the frequencies of electron vertical transitions, $\Theta$ is the Heaviside function.

From Eqs. (9) and (10) one can see that the real part of the polarization operator, as the function of the frequency $\omega$, has logarithmic singularities at the boundaries of the regions, on the plane $(q,\omega)$, in which the collisionless damping of plasma waves is absent. The boundaries of these regions are derivable from the formula (10). The imaginary part of the polarization operator is equal to zero in the spectral windows for plasma waves that are limited by parabolas on the plane $(q,\omega)$.

It follows from the equation (10) that each sub-zone in the electron energy spectrum is related to a branch of plasma waves that propagate along the tube axis. In particular, for the branch with the number $m=0$, the frequencies of the transitions (12) are equal to zero and the dispersion equation (4) is simplified. Let us present the solution of this equation in the absence of magnetic field in the ultra-quantum limit, when only the lowermost sub-zone $m'=0$ is filled up. In this case, there is a parabolic spectral window which is limited by the parabola $\omega = q\upsilon_0-\omega_q$ and by the axis $q$ on the plane $(q,\omega)$, the region is above the parabola $\omega = q\upsilon_0+\omega_q$ and to the right of the parabola $\omega = -q\upsilon_0+\omega_q$, in which the collisionless damping of waves is absent. The analysis of the equation (4) indicates that in this case there are no solutions for the equation (4) in the parabolic spectral window. They do exist above the parabola $\omega = q\upsilon_0+\omega_q$ and in the region of the collisionless damping. Above the parabola $\omega = q\upsilon_0+\omega_q$, we have the following solution:

$$\omega_0(q) = q\upsilon_0 + \omega_q \mathrm{cth} \frac{2\pi^2 q}{m_* \upsilon_0(q)}. \tag{13}$$

In the long-wavelength approximation, $qa \ll 1$, we shall use the expansions [10]:

$$I_0(x)K_0(x) \approx \ln\frac{2}{xe^\gamma},$$

$$I_1(x)K_1(x) \approx \frac{1}{2}\left[1 + \frac{x^2}{2}\ln\frac{x}{2}\right], \tag{14}$$

$$I_m(x)K_m(x) \approx \frac{1}{2m}\left[1 - \frac{\left(\frac{x}{2}\right)^2}{m-1}\right] \quad (m = 2, 3, ...),$$

where $x \ll 1$, $\gamma = 0{,}577...$ is the Euler number. In this approximation and using the formula (13) we obtain the plasmon spectrum:

$$\omega_0(q) = \frac{\bar{e}^2}{\pi} q \ln\frac{2}{aqe^\gamma}. \tag{15}$$

The dispersion of this wave is normal. In the region $qa \gg 1$ the dispersion curve (13) approaches the parabola $\omega = q\upsilon_0 + \omega_q$ from above.

The plasma wave collisionless damping region is situated on the plane $(q, \omega)$ between the parabolas $\omega = q\upsilon_0 + \omega_q$, $\omega = q\upsilon_0 - \omega_q$ and $\omega = -q\upsilon_0 + \omega_q$. In this region the solution of the dispersion equation (4) in the case under consideration has the form:

$$\omega_0(q) = q\upsilon_0 + \omega_q \mathrm{th} \frac{2\pi^2 q}{m_* \upsilon_0(q)}. \tag{16}$$

Hence in the long-wavelength limit we obtain:

$$\omega_0(q) = q\upsilon_0 + \frac{\pi q^3}{4m_*^2 \bar{e}^2 \ln\dfrac{2}{aqe^\gamma}}.$$

In the limit $qa \gg 1$ the dispersion curve (16) approaches the parabola $\omega = q\upsilon_0 + \omega_q$ from below. It corresponds to the damping plasma wave.

In the absence of magnetic field in the case of long wavelengths $q\upsilon \ll \omega$ ($\upsilon$ – the Fermi electron velocity) we obtain the following equation from the formula (9):

$$\mathrm{Re}\, P_0(q, \omega) = \frac{qL}{\pi\omega} \sum_{m'} \Theta(\mu_0 - \varepsilon_{m'}), \tag{17}$$

where $\varepsilon_{m'} = \varepsilon_0 m'^2$ is the sub-zone boundary with the number $m'$. In case of a large number of filled sub-zones, we use the Poisson formula to calculate the sum:

$$\sum_{m=-\infty}^{\infty} \varphi(m) = \sum_{l=-\infty}^{\infty} \int_{-\infty}^{\infty} dx\, \varphi(x) e^{2\pi i l x}. \tag{18}$$

Then for the mode $m = 0$ in the limit $qa \ll 1$ we obtain:

$$\omega_0(q) = \frac{4\bar{e}^2}{\pi} q \ln\frac{2}{qae^\gamma} \sqrt{\frac{\mu_0}{\varepsilon_0}} \left(1 + \frac{1}{\pi}\sqrt{\frac{\varepsilon_0}{\mu_0}} \sum_{l=1}^{\infty} \frac{1}{l} \sin 2\pi l \sqrt{\frac{\mu_0}{\varepsilon_0}}\right). \tag{19}$$

The wave frequency undergoes quantum oscillations of the de Haas-van Alfven type with the change of the Fermi energy due to the crossing of the sub-zone boundaries by the Fermi level. The Fermi energy is associated with the linear density $n = N/L$ of electrons via the relationship:

$$n = \frac{2}{\pi}\sqrt{2m_*\varepsilon_0}\sum_m \left(\frac{\mu_0}{\varepsilon_0} - m^2\right)^{1/2}.$$

The period of oscillations due to varying of $\sqrt{\mu_0}$ is equal to $1/\sqrt{2m_*}a$. It is determined by the electron effective mass and by the tube radius. The relative oscillation amplitude $\sim \sqrt{\frac{\varepsilon_0}{\mu_0}}$ is small for $\mu_0 \gg \varepsilon_0$.

Let us consider the dispersion of modes with the numbers $m > 0$. In the long-wavelength limit we obtain from the formula (9) that

$$\operatorname{Re} P_m = \frac{2m_* L}{\pi}\sum_{m'}\frac{\upsilon_{m'} - \upsilon_{m'+m}}{\omega - \Omega_{mm'}}, \tag{20}$$

where the summation is done over those values of $m'$, at which the sub-root expressions in (11) are positive. The sum in Eq. (20) is calculated according to the Poisson formula (18). In the dispersion equation (34) we still assume that $qa \ll 1$. In addition, we shall restrict ourselves to the case of high frequencies that satisfies the inequality:

$$\omega \gg \varepsilon_0\left[\left(m + \sqrt{\frac{\mu_0}{\varepsilon_0}}\right)^2 - \frac{\mu_0}{\varepsilon_0}\right].$$

Here $\left[\sqrt{\frac{\mu_0}{\varepsilon_0}}\right]$ is the number of filled sub-zones. Then from the formula (20) we obtain:

$$\operatorname{Re} P_m = \frac{2\sqrt{2m_*\varepsilon_0}\,m^2 \mu_0 L}{(\omega^2 - \varepsilon_m^2)}\left[1 + \frac{2}{\pi}\sqrt{\frac{\varepsilon_0}{\mu_0}}\sum_{l=1}^{\infty}\frac{1}{l}J_1\left(2\pi l\sqrt{\frac{\mu_0}{\varepsilon_0}}\right)\right], \tag{21}$$

where $J_1$ is the Bessel function. In this case, the solutions of the equation (4) have the form:

$$\omega_1^2(q) = \varepsilon_1^2 + 2\bar{e}^2\sqrt{2m_*\varepsilon_0}\,\mu_0\left[1 + \frac{1}{2}(aq)^2 \ln\frac{aq}{2}\right]\left[1 + \frac{2}{\pi}\sqrt{\frac{\varepsilon_0}{\mu_0}}\sum_{l=1}^{\infty}\frac{1}{l}J_1\left(2\pi l\sqrt{\frac{\mu_0}{\varepsilon_0}}\right)\right],$$

$$\omega_m^2(q) = \varepsilon_m^2 + 2\bar{e}^2\sqrt{2m_*\varepsilon_0}\,m\mu_0\left[1 - \frac{(aq)^2}{4(m-1)}\right]\left[1 + \frac{2}{\pi}\sqrt{\frac{\varepsilon_0}{\mu_0}}\sum_{l=1}^{\infty}\frac{1}{l}J_1\left(2\pi l\sqrt{\frac{\mu_0}{\varepsilon_0}}\right)\right] \tag{22}$$

$$(m = 2, 3, ...).$$

The cut-off frequencies for the waves of the spectrum (22) are equal to:

$$\omega_m^2(0) = \varepsilon_m^2 + 2\bar{e}^2\sqrt{2m_*\varepsilon_0}\,m\mu_0\left[1 + \frac{2}{\pi}\sqrt{\frac{\varepsilon_0}{\mu_0}}\sum_{l=1}^{\infty}\frac{1}{l}J_1\left(2\pi l\sqrt{\frac{\mu_0}{\varepsilon_0}}\right)\right] \tag{23}$$

$$(m = 1, 2, ...).$$

The dispersion of these waves is anomalous. Their frequencies undergo the oscillations considered above associated with the crossing of the sub-zone boundaries by the Fermi level.

## II.  Non-degenerate electron gas

In this Section we make use of the Boltzmann distribution function to compute the polarization operator (9), (10) in the absence of magnetic field. The real and imaginary parts of the polarization operator of the Boltzmann electron gas are as follows:

$$\operatorname{Re} P_m(q,\omega) = \frac{N}{q}\sqrt{\frac{m_*\beta}{2}} \left\langle F(x^-_{mm'}) - F(x^+_{mm'}) \right\rangle, \qquad (24)$$

$$\operatorname{Im} P_m(q,\omega) = \frac{N}{|q|}\sqrt{\frac{\pi m_*\beta}{2}} \left\langle \exp\left(-\beta\frac{k^{+2}_{mm'}}{2m_*}\right) - \exp\left(-\beta\frac{k^{-2}_{mm'}}{2m_*}\right) \right\rangle, \qquad (25)$$

where

$$\langle ... \rangle = \frac{\sum_{m'} e^{-\beta\varepsilon_0 m'^2} ...}{\sum_{m'} e^{-\beta\varepsilon_0 m'^2}},$$

$$F(x) = \frac{1}{\sqrt{\pi}} \text{P.}\int_{-\infty}^{\infty} dy \frac{e^{-y^2}}{x-y},$$

$$x^{\pm}_{mm'} = \frac{1}{q}\sqrt{\frac{m_*\beta}{2}}(\omega_\pm \pm \Omega^{\pm}_{mm'}), \quad \Omega^{\pm}_{mm'} = 2m\varepsilon_0\left(\frac{m}{2}\mp m'\right), \quad k^{\pm}_{mm'} = \sqrt{\frac{2m_*}{\beta}} x^{\pm}_{mm'}, \quad \omega_\pm = \omega \pm \omega_q,$$

$\beta$ is the inverse temperature. The number of electrons $N$, entering into (24) and (25), is connected with the chemical potential $\mu$ via the relation:

$$N = \sqrt{\frac{2m_*}{\pi\beta}} L e^{\beta\mu} \sum_{m=-\infty}^{\infty} e^{-\beta\varepsilon_0 m^2} = \sqrt{\frac{2m_*}{\varepsilon_0}} \frac{L}{\beta} e^{\beta\mu} \sum_{l=-\infty}^{\infty} \exp\left(-\frac{\pi^2 l^2}{\beta\varepsilon_0}\right).$$

Here we have used the formula [11]:

$$\sum_{m=-\infty}^{\infty} \exp\left[-(m+b)^2 x\right] = \sqrt{\frac{\pi}{x}} \sum_{l=-\infty}^{\infty} \exp\left(-\frac{\pi^2 l^2}{x}\right)\cos 2\pi l b. \qquad (26)$$

To calculate the spectrum and the damping of the mode with the number $m=0$ we use the following expressions:

$$\operatorname{Re} P_0(q,\omega) = \frac{N}{q}\sqrt{\frac{m_*\beta}{2}} \left[ F\left(\sqrt{\frac{m_*\beta}{2}}\frac{\omega_-}{q}\right) - F\left(\sqrt{\frac{m_*\beta}{2}}\frac{\omega_+}{q}\right) \right],$$

$$\operatorname{Im} P_0(q,\omega) = -\frac{N}{|q|}\sqrt{2\pi m_*\beta} \, \operatorname{sh}\frac{\beta\omega}{2} \exp\left(-\frac{\beta m_*\omega^2}{2q^2} - \frac{\beta q^2}{8m_*}\right). \qquad (27)$$

In the long-wavelength limit $qr_D \ll 1$ ($r_D$ is the Debye shielding radius), we obtain from Eq. (27) the following expression:

$$\operatorname{Re} P_0(q,\omega) = \frac{Nq^2}{m_*\omega^2},$$

$$\operatorname{Im} P_0(q,\omega) = -\frac{N}{|q|}\sqrt{\frac{\pi m_*\beta}{2}}\beta\omega\exp\left(-\frac{\beta m_*\omega^2}{2q^2}\right). \tag{28}$$

Then from the dispersion equation in the case of $qa \ll 1$, we derive the plasma wave spectrum

$$\omega_0^2(q) = \frac{2\bar{e}^2 n}{m_*}q^2 \ln\frac{2}{aqe^{\gamma}}. \tag{29}$$

The decrement of the damping of this wave

$$\gamma(q) = \frac{\operatorname{Im} P(q,\omega(q))}{\dfrac{\partial}{\partial \omega(q)}\operatorname{Re} P(q,\omega(q))} \tag{30}$$

is equal to

$$\gamma_0(q) = \sqrt{\frac{\pi}{8}}\left(\frac{m_*\beta}{q^2}\right)^{3/2}\omega_0^4(q)\exp\left[-\frac{\beta m_*}{2}\left(\frac{\omega_0(q)}{q}\right)^2\right]. \tag{31}$$

The dispersion of the wave with the spectrum (29) is normal, the damping decrement (31) diminishes exponentially at $q \to 0$.

The real part of the polarization operator (24) for $|m| > 0$ within the long-wavelength limit is:

$$\operatorname{Re} P_m = \frac{N}{2\varepsilon_0}\left\langle \left\{\left[m' - \frac{\omega}{2m\varepsilon_0}\right]^2 - \frac{m^2}{4}\right\}^{-1}\right\rangle. \tag{32}$$

The sums $\sum\limits_{m'}$ in the case of $\beta\varepsilon_0 \ll 1$ can be replaced by integrals. Therefore

$$\operatorname{Re} P_m = -\frac{N}{2m}\sqrt{\frac{\beta}{\varepsilon_0}}\left\{\operatorname{F}\left(\sqrt{\beta\varepsilon_0}\left[\frac{\omega}{2m\varepsilon_0}+\frac{m}{2}\right]\right) - \operatorname{F}\left(\sqrt{\beta\varepsilon_0}\left[\frac{\omega}{2m\varepsilon_0}-\frac{m}{2}\right]\right)\right\}. \tag{33}$$

In the case of the high frequencies that satisfy the inequality

$$\omega \gg 2m\varepsilon_0\left(\frac{m}{2}+\frac{1}{\sqrt{\beta\varepsilon_0}}\right),$$

one can use the asymptote of the function $\operatorname{F}(x) \approx x^{-1}$ at $x \gg 1$. Then the expression (33) is approximated by

$$\operatorname{Re} P_m = \frac{2N\varepsilon_0 m^2}{\omega^2 - \varepsilon_m^2}. \tag{34}$$

By substituting this expression into the dispersion Eq. (4) at $qa \ll 1$, we obtain:

$$\omega_1^2(q) = \varepsilon_0^2 + 2\bar{e}^2\varepsilon_0 n\left[1 + \frac{(aq)^2}{2}\ln\frac{aq}{2}\right], \tag{35}$$

$$\omega_m^2(q) = \varepsilon_m^2 + 2\overline{e}^2 \varepsilon_0 n |m| \left[1 - \frac{a^2 q^2}{4(|m|-1)}\right] \qquad (m = \pm 2, \pm 3, ...). \tag{36}$$

The wave dispersion with the spectrum (35), (36) is anomalous. The cut-off frequencies in the spectrum of these waves are

$$\omega_m^2(0) = \varepsilon_m^2 + 2\overline{e}^2 \varepsilon_0 n |m|. \tag{37}$$

To compute the decrement of the damping of the modes with the numbers $|m| > 0$, we shall make use of the formulae (30) and (34) and of the imaginary part of the polarization operator (25) that is equal in the long-wavelength limit to:

$$\operatorname{Im} P_m = -\frac{N}{|m|} \sqrt{\frac{\pi \beta}{\varepsilon_0}} \operatorname{sh} \frac{\beta \omega}{2}. \tag{38}$$

Then the decrement of the damping of the waves with the spectrum (35), (36) is as follows:

$$\gamma_1(q) = \frac{\overline{e}^4 \varepsilon_0 n^2}{\omega_1(q)} \sqrt{\frac{\pi \beta}{\varepsilon_0}} \operatorname{sh} \frac{\beta \omega_1(q)}{2} \left[1 + \frac{(aq)^2}{2} \ln \frac{aq}{2}\right]^2, \tag{39}$$

$$\gamma_m(q) = \frac{\overline{e}^4 \varepsilon_0 n^2}{|m| \omega_m(q)} \sqrt{\frac{\pi \beta}{\varepsilon_0}} \operatorname{sh} \frac{\beta \omega_m(q)}{2} \left[1 - \frac{a^2 q^2}{4(|m|-1)}\right]^2 \qquad (m = \pm 2, \pm 3, ...). \tag{40}$$

The ratio $\gamma_m/\omega_m$ decreases with increasing mode number in proportion to $|m|^{-3}$.

### III. Magnetoplasma waves

In the magnetic field, which is parallel to the cylinder axis, the energy of electron is given by Eq. (7), while the polarization operator of degenerate electron gas is derived in Eq.(9), (10).

Let us obtain the solution of the dispersion equation (4) in the ultra-quantum limit, when only two sub-zones are filled with the numbers $(m,\sigma) = (0,\pm) = 0^{\pm}$. In this case, there are lobe and triangular spectral windows in the vector-frequency wave plane besides the parabolic spectral window which is found between the parabola $\omega = q\upsilon_0^+ - \omega_q$ and the axis $q$, and also besides the collisionless wave damping region above the parabola $\omega = q\upsilon_0^- + \omega_q$ and to the right of the parabola $\omega = -q\upsilon_0^- + \omega_q$. The lobe spectral window is confined within the parabolas $\omega = q\upsilon_0^+ + \omega_q$ and $\omega = q\upsilon_0^- - \omega_q$, and the triangular spectral window is limited by the parabolas $\omega = -q\upsilon_0^+ + \omega_q$, $\omega = q\upsilon_0^- - \omega_q$ and the axis $q$. The coordinates of the lobe window uppermost boundary in the plane $(q, \omega)$ are $\left(m_*\left(\upsilon_0^- - \upsilon_0^+\right), \frac{m_*}{2}\left(\upsilon_0^{-2} - \upsilon_0^{+2}\right)\right)$. If $q < m_*\left(\upsilon_0^- - \upsilon_0^+\right)$, the solutions for the equation (4) exist in the lobe spectral window above the parabola $\omega = q\upsilon_0^- + \omega_q$ and in the regions of collisionless damping. The dispersion law of magnetoplasma wave in the lobe spectral window is

$$\omega_0(q) = q\frac{\upsilon_0^+ + \upsilon_0^-}{2} + \omega_q \text{cth} \frac{2\pi^2 q}{m_* \upsilon_0(q)} - \left[ \frac{q^2}{4}\left(\upsilon_0^+ - \upsilon_0^-\right)^2 + \frac{\omega_q^2}{\text{sh}^2 \frac{2\pi^2 q}{m_* \upsilon_0(q)}} \right]^{1/2}. \quad (41)$$

At small values of $q$ the spectrum of this wave is linear. With increasing $q$, the dispersion curve (41) comes closer to the lobe boundary $\omega = q\upsilon_0^+ + \omega_q$. The appearance of the quantum number of the magnetic flux $\Phi/\Phi_0$ in the formula (7) affects the wave spectrum considered in Section 1. In particular, the formula (19), in the presence of magnetic field and taking into account Eq.(26), takes the following form:

$$\omega_0(q) = \frac{4\overline{e}^2}{\pi} q \ln \frac{2}{qae^\gamma} \sqrt{\frac{\mu_0}{\varepsilon_0}} \left( 1 + \frac{1}{\pi}\sqrt{\frac{\varepsilon_0}{\mu_0}} \sum_{l=1}^\infty \frac{1}{l} \sin 2\pi l \sqrt{\frac{\mu_0}{\varepsilon_0}} \cos 2\pi l \frac{\Phi}{\Phi_0} \right). \quad (42)$$

This formula does not account for the spin splitting of the levels. Besides the oscillations of the de Haas-Van Alfven type, considered in Section 1, which appear due to the change of $\mu_0$, the wave frequency sustains oscillations of the Aharonov-Bohm type which appear when the magnetic field is changing. The period of these oscillations is equal to one quantum of the magnetic flux $\Phi_0$. In the magnetic field, the linear electron density $n$ is connected with the chemical potential $\mu_0$ via the relation:

$$n = \frac{2}{\pi}\sqrt{2m_*\varepsilon_0} \sum_m \left[ \frac{\mu_0}{\varepsilon_0} - \left(m + \frac{\Phi}{\Phi_0}\right)^2 \right]^{1/2}.$$

The frequencies of modes with the numbers $|m| > 0$ also undergo the Aharonov-Bohm type oscillations. In the formulae (22) and (23), under the sign of the sum $\sum_l$, it appears the factor $\cos 2\pi l \frac{\Phi}{\Phi_0}$.

## Conclusions

The energy spectrum (2) of electron on the semiconductor nanotube surface is a set of one-dimensional sub-zones, the positions of the boundaries of which are not equidistant. As a result, the density of states (3) oscillates, with a period $\left(\sqrt{2m_*}a\right)^{-1}$, when changing $\sqrt{\varepsilon}$. This accounts for the oscillations of plasma wave frequencies in the degenerate electron gas with changing $\sqrt{\mu_0}$. These oscillations resemble the de Haas-Van Alfven oscillations of electron gas magnetization that emerge with changing magnetic field strength [12]. This difference is attributed to the existence of non-equi-distant boundaries of the spectrum sub-zones (2). The cause of the oscillations is a jump of the density of states when the Fermi level traverses a sub-zone boundary. The measurement of the period of the oscillations allows to calculate $\sqrt{m_*}a$. In order to observe these oscillations it is necessary to be able to change the Fermi level of electrons on the tube, as it is done in the two-dimensional electron gas [13]. While measuring the plasmon frequencies on the semiconductor tubes with different values of $m_*, a, \mu_0$, a spread of the frequency values should be expected, caused by the oscillations.

In a magnetic field, which is parallel to the cylinder axis, the Aharonov-Bohm oscillations appearing with the changing magnetic field will be superposed to the magnetoplasma wave frequency oscillations of the de Haas-Van Alfven type. The reason for the former oscillations differs from the de Haas-Van Alfven type. Their period does not depend on $\mu_0$ and is equal to one quantum of the magnetic flux $\Phi_0$. It is related to the area, $\pi a^2$, occupied by the projection of the orbit of electron to the plane $z=0$, and it does not depend on the energy of the electron.

The oscillations described here can be observed in experiments measuring the cross-section of scattering of light and electrons by plasma waves on the semiconductor nanotube surface.

We would like to thank T. Rashba for help with preparation of the manuscript.